\documentclass[a4paper]{article}

\usepackage{INTERSPEECH2020}
\usepackage{mathrsfs}
\usepackage{diagbox}
\usepackage{multirow}
\usepackage{subfigure}
\usepackage{threeparttable}
\usepackage{algorithm}
\usepackage{algorithmic}
\usepackage{bbding}

\title{FSR: Accelerating the Inference Process of Transducer-Based Models \\ by Applying Fast-Skip Regularization}
\name{Zhengkun Tian$^{1,2}$, Jiangyan Yi$^{1}$, Ye Bai$^{1,2}$, Jianhua Tao$^{1,2,3}$, Shuai Zhang$^{1,2}$, Zhengqi Wen$^{1}$}
\address{
    $^1$NLPR, Institute of Automation, Chinese Academy of Sciences, Beijing, China\\
    $^2$School of Artificial Intelligence, University of Chinese Academy of Sciences, Beijing, China\\
    $^3$ CAS Center for Excellence in Brain Science and Intelligence Technology, Beijing, China}
\email{\{zhengkun.tian, jiangyan.yi, ye.bai, jhtao, shuai.zhang, zqwen\}@nlpr.ia.ac.cn}

\begin{document}

\maketitle
\begin{abstract}
Transducer-based models, such as RNN-Transducer and transformer-transducer, have achieved great success in speech recognition. A typical transducer model decodes the output sequence conditioned on the current acoustic state and previously predicted tokens step by step. Statistically, The number of blank tokens in the prediction results accounts for nearly 90\% of all tokens. It takes a lot of computation and time to predict the blank tokens, but only the non-blank tokens will appear in the final output sequence. Therefore, we propose a method named fast-skip regularization, which tries to align the blank position predicted by a transducer with that predicted by a CTC model. During the inference, the transducer model can predict the blank tokens in advance by a simple CTC project layer without many complicated forward calculations of the transducer decoder and then skip them, which will reduce the computation and improve the inference speed greatly. All experiments are conducted on a public Chinese mandarin dataset AISHELL-1. The results show that the fast-skip regularization can indeed help the transducer model learn the blank position alignments. Besides, the inference with fast-skip can be speeded up nearly 4 times with only a little performance degradation.

\end{abstract}
\noindent\textbf{Index Terms}: Transducer-based Models, Speech Recognition, Fast-Skip Regularization, CTC model

\section{Introduction}

% The end-to-end models, such as attention-based encoder-decoder model \cite{bahdanau2014neural, chorowski2015attention, chan2016listen, kim2017joint, dong2018speech}, CTC-based model and transducer-based model, have replaced the hybrid GMM-HMM model as the mainstream models for speech recognition.
Transducer-based models \cite{graves2012sequence, rao2017exploring, bagby2018efficient, sainath2019two}, such as the RNN-Transducer (RNN-T) \cite{li2019improving, graves2013speech, he2019streaming} and Transformer-Transducer (T-T) \cite{Tian2019, zhang2020transformer, yeh2019transformer}, have been widely applied for speech recognition. Compared with the attention-based encoder-decoder models \cite{bahdanau2014neural, chorowski2015attention, chan2016listen, kim2017joint, dong2018speech}, the transducer-based models predict a new token conditioned on the previous output tokens and the current encoded frame rather than the whole acoustic encoded sequence, which make it can be directly applied for streaming speech recognition. Different from the CTC model \cite{graves2006connectionist, hannun2014deep, li2019jasper}, transducer-based models can model the linguistic dependencies between the output tokens.

\begin{figure}[t]
    \centering
    \includegraphics[width=0.6\linewidth]{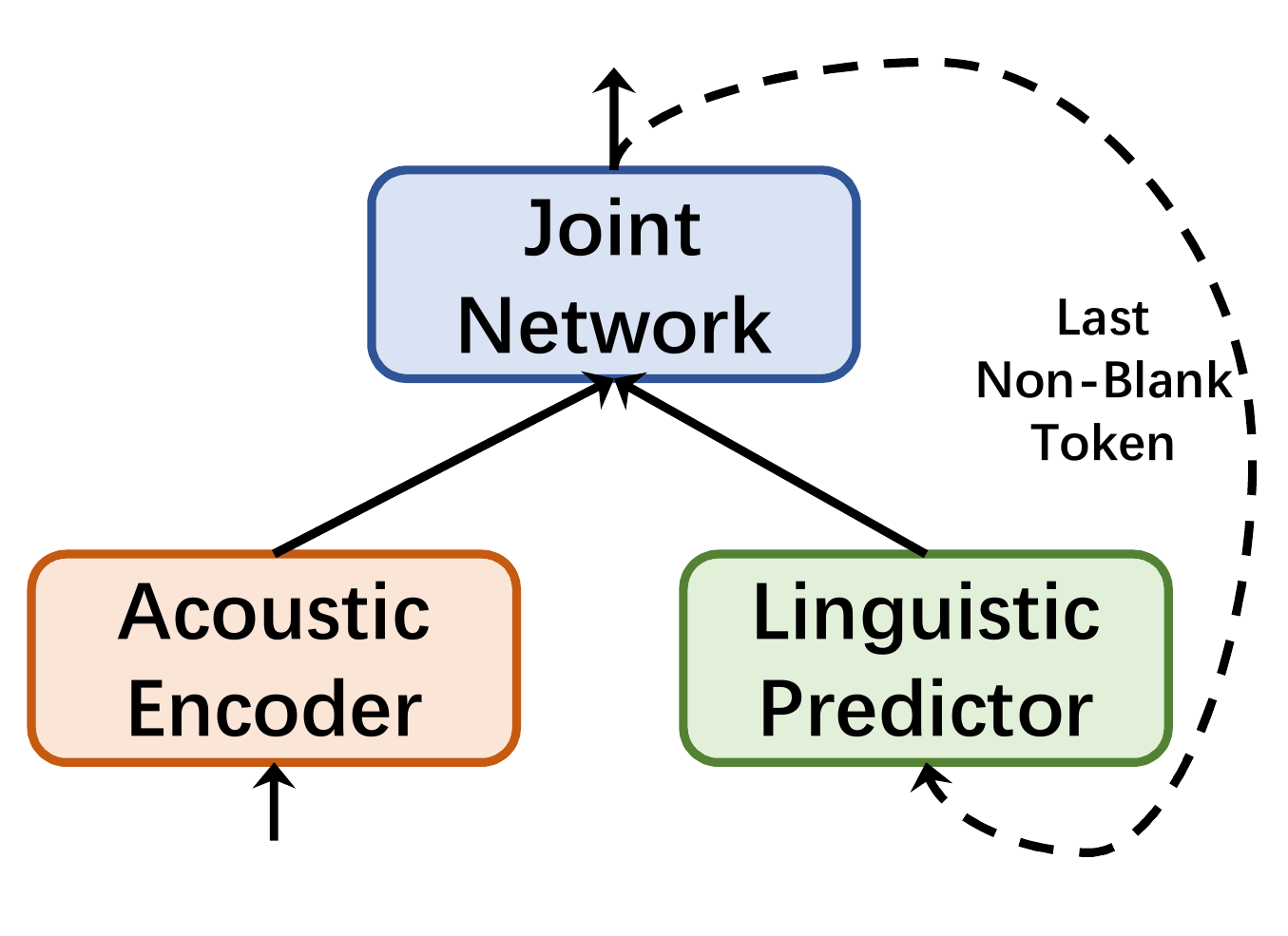}
    \vspace{-10pt}
    \caption{The Architecture of Transducer-Based Models}
    \vspace{-20pt}
    \label{fig:transducer}
\end{figure}

A typical transducer-based model consists of three components, an acoustic encoder (also named transcription network), a linguistic predictor, and a joint network, as shown in Fig.\ref{fig:transducer}. As the same as the CTC models, the transducer models adopt the forward-backward algorithm to optimize all possible alignment paths and decode the output sequence frame by frame. This characteristic affects the inference speed of the transducer models from two aspects. On the one hand, the frame-wise operation forces the transducer to pass through the transducer decoder (including the linguistic predictor and the joint network) at each step of inference, which leads to a lot of calculations. On the other hand, due to the fact that the frame-level output probability lattice contains a lot of alignment paths with the same prefixes, the transducer models have to merge the probability of these duplicated paths. To our knowledge, there are four kinds of methods to address these two problems. Firstly, the neural transducer \cite{jaitly2015neural} and sync-transformer \cite{tian2019synchronous} split the acoustic encoded states into some fixed-length chunk and predict the output sequence chunk by chunk, which has made a significant change in the structure of the transducer model. Secondly, the neural aligner \cite{sak2017recurrent, dong2018extending, dong2019self} forces the model to decode at most one token based on one acoustic frame, which simplifies the output probability lattice of both the training and inference but it still adopts frame-by-frame inference. Thirdly, \cite{kim2020accelerating} also directly applied the one-step constraint to the inference of transducer-based model, which is a simple and limited method. Finally, a recently proposed model named State-Less RNN-T \cite{ghodsi2020rnn, zhang2021tiny} utilizes a simple embedding layer instead of the traditional multiple RNN layer as the linguistic predictor, which improves the inference speed by reducing the computation of predictor. However, this method also makes it difficult for the transducer to capture the long-range linguistic dependencies.

We notice that the transducer model will produce a lot of blank tokens during inference. The blank token stands for silence or duplicated token, which will not appear in the final predicted result. The prediction of each token must depend on the forward calculation of the transducer decoder. Furthermore, we find the predicted blank tokens account for over 90\% of all the tokens. Based on this observation, we assume that the inference process can be speeded up by predicting the blank tokens in advance with some simple computation and then skipping them. In this paper, we propose a novel method named fast-skip regularization (FSR) to make the transducer model depend on a CTC project layer to predict the blank tokens in advance. The FSR introduces a CTC module into the transducer model. And it forces the transducer model to align the blank position with that predicted by the CTC project layer. During the inference, the transducer model can skip the blank token based on the prediction of the CTC project layer, which makes the output probability lattice be sparse and improve the inference speed greatly. All the experiments are conducted on a public Chinese mandarin dataset AISHELL-1. The results show that the FSR can align the blank position predicted by the CTC and transducer accurately. The inference process can be accelerated up to 4 times with only a little performance degradation.

The remainder of this paper is organized as follows. We introduce the details of our methods in Section 2. Our experimental setup and results will be presented in Section 3. The conclusions and future works will be given in Section 4.

\section{Fast-Skip Regularization for Transducer-based Models}
\subsection{Transducer}
Transducer models can transcribe the acoustic feature sequence into the target sequence directly. As the same as the CTC model, the transducer introduces a blank token $\varnothing$ to represent the duplicated tokens and silence. Given an input sequence $\bm{x}=(x_1, x_2, ..., x_{T})$ with length $T$ and the target sequence $\bm{y}=(y_1,y_2,...,y_U)$ with length $U$, the transducer aims to maximum the log-probability $P(\bm{y}|\bm{x})$:
\begin{equation}
    \mathcal{L}_{transducer}=-\text{ln}P(\bm{y}|\bm{x})=-\text{ln}\sum_{\pi \in \mathcal{B}^{-1}(\bm{y})} P(\pi|\bm{x})
\end{equation}
where $\pi$ indicates any one alignment path and $\mathcal{B}_{-1}(\bm{y})$ represents the space which contains all possible alignment path, as shown in Fig.2(a). It is very challenging to calculate $P(\bm{y}|\bm{x})$ by enumerating all possible alignment paths. Therefore, the transducer introduces an efficient forward-backward algorithm to optimize all possible alignment paths \cite{graves2012sequence}.

The \textit{forward variable} $\alpha(t,u)$ means the sum of the probability of all the possible paths, which begins with the start token $y_0$ ($=\varnothing$) and ends with $y_u$ at the $t$-th frame. Given the $t$-th frame and the predicted token sequence $y_{0:u-1}$, the probabilities of predicting $\varnothing$ and $y_{u}$ are represented as $\varnothing(t, u)$ and $y(t, u)$ respectively. For all $1 \leq t \leq T$ and $ 0 \leq u \leq U$, the forward variables can be calculated recursively using

\begin{equation}
    \alpha(t,u) = \alpha(t-1,u)\varnothing (t, u) + \alpha(t,u-1)y(t, u-1)
\end{equation}
And all paths begin with $y_0$ ($=\varnothing$), it means $\alpha(1,0)=1$. The conditional probability $P(\bm{y}|\bm{x})$ can be expressed by the forward variable at the terminal node.
\begin{equation}
    P(\bm{y}|\bm{x}) = \alpha(T,U)\phi(T, U)
\end{equation}

Similarly, the \textit{backward variable} $\beta(t,u)$ means the sum of the probabilities of all possible paths, which begin with $y_u$ at $t$-th frame and end with $y_{U}$(=$\varnothing$) in the last frame. The backward variables can be expressed as

\begin{equation}
    \beta(t,u) = \beta(t+1,u)\varnothing(t, u) + \beta(t,u+1)y(t, u)
\end{equation}
where the initial condition $\beta(T,U)=\varnothing(T, U)$.

Given an input feature sequence $\bm{x}$ and a target sequence $\bm{y}$, the probability $p(\bm{y}|\bm{x})$ is equal to the sum of $\alpha(t,u)\beta(t,u)$ over any top-left to bottom-right diagonal through the nodes. That is, $\forall n: 1 \leq n \leq U+T$

\begin{equation}
   P(\bm{y}|\bm{x})=\sum_{(t, u):t+u=n}\alpha(t,u)\beta(t,u)
\end{equation}

\subsection{Fast-Skip Regularization}
We try to make the transducer model able to predict the blank tokens $\varnothing$ in advance without depending on the forward of the transducer decoder. Therefore, we propose a method named Fast-Skip Regularization (FSR), which forces the transducer model to align the predicted blank position with the blank position of the CTC model.

Inspired by the FastEmit \cite{yu2020fastemit, li2020better, narayanan2020cascaded}, we utilize FSR to adjust the probability of possible alignment paths on the output lattice of the transducer model. We define a variable $\mathcal{A}_{t,u}$ to represent all the possible alignment paths that pass thought the \textit{node} $(t,u)$, as the blue node shown in Fig.2(a). The conditional probability $P(\mathcal{A}_{t,u}|\bm{x})$ can be decomposed into two terms:
\begin{equation}
    \begin{aligned}
        &P(\mathcal{A}_{t,u}|\bm{x})=\alpha(t,u)\beta(t,u)\\
        &=P(\mathcal{A}^{nb}_{t,u}|\bm{x})+P(\mathcal{A}^{b}_{t,u}|\bm{x})=\\
        &\underbrace{\alpha(t,u)y(t,u)\beta(t,u+1)}_{\text{predict non-blank token}}
        +\underbrace{\alpha(t,u)\varnothing(t,u)\beta(t+1,u)}_{\text{predict blank token}}\\
    \end{aligned}
\end{equation}
where the first term $P(\mathcal{A}^{nb}_{t,u}|\bm{x})$ indicates the probability of all the alignment path transferred from the blue cirle to the green circle (vertical direction) and the second term $P(\mathcal{A}^{b}_{t,u}|\bm{x})$ means the probability of all the alignment paths that transferred from the bule circle to the orange circle (horizontal direction).

\begin{figure}[t]
    \centering
    \includegraphics[width=\linewidth]{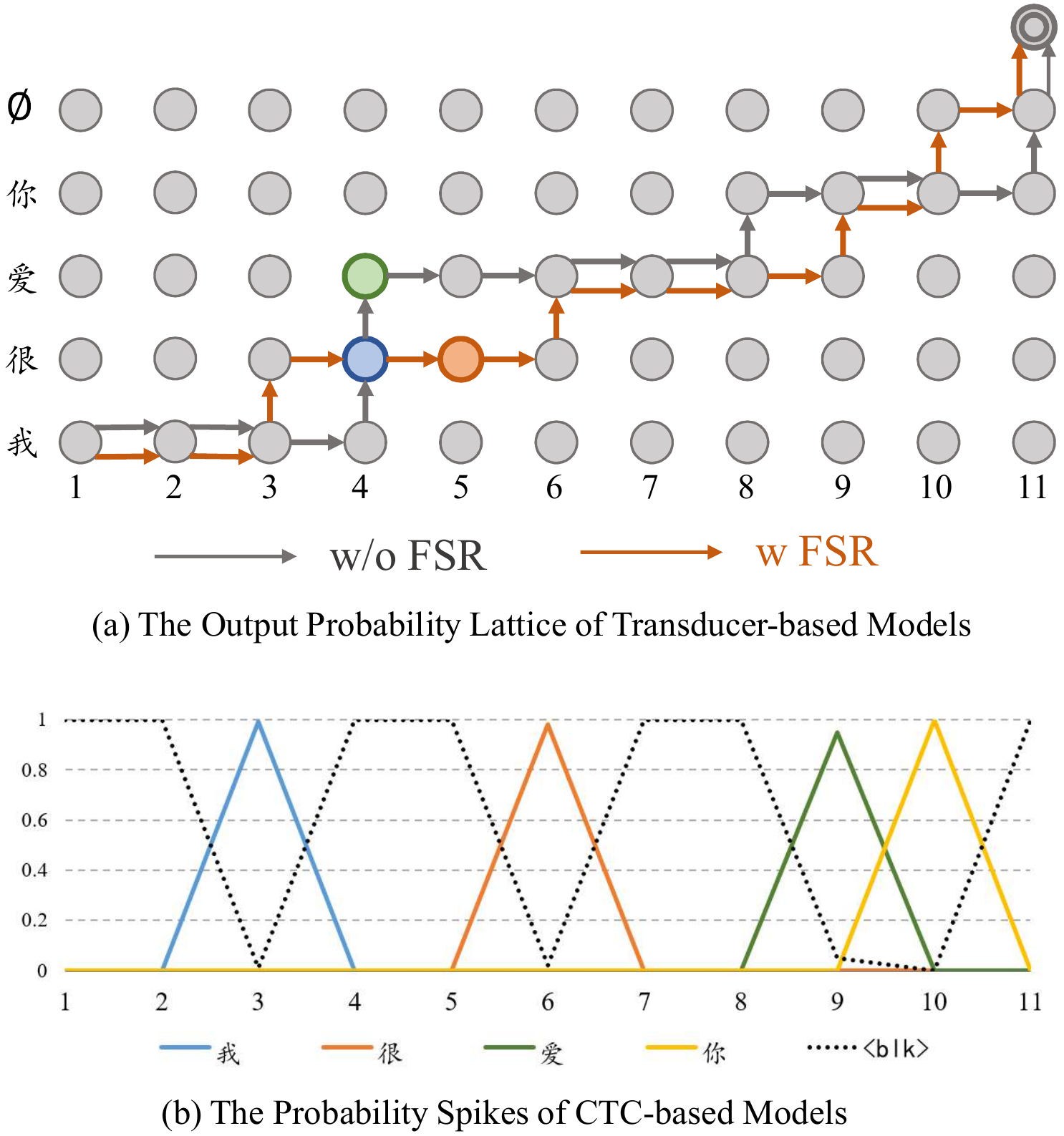}
    \vspace{-10pt}
    \caption{Illustration of Fast-Skip Regularization. (a) indicates the output probability lattice of the transducer-based model. The gray transfer path means the alignment sequence without FSR. The orange transfer path means our expected alignment sequence with FSR. The symbol $\varnothing$ means the blank token. (b) shows the spikes graph generate by the CTC project layer. The FSR will force the vertical non-blank transfer of the transducers to align with the non-blank spikes generated by the CTC models. And the horizontal transfers are consistent with the blank positions of the CTC model.}
    \vspace{-15pt}
    \label{fig:align}
\end{figure}

Based on the observation that the transducer loss aims to maximum the log-probability regardless of their emission position \cite{yu2020fastemit}, we try to make the blank token positions predicted by these two models be consistent. We introduce a CTC module into the transducer model by putting a project linear layer at the top of the acoustic encoder. Afterwards, we define the blank probability predicted by the CTC module based on the $t$-th frame as $\mathcal{C}_t^{b}$, as shown by the black dot line in Fig.2(b). Similarly, $\mathcal{C}_t^{nb}$, as shown by the colorful spikes in Fig.2(b), indicates the non-blank probability. And there is the following equation relationship between them: $\mathcal{C}_t^{nb}=1-\mathcal{C}_t^{b}$. We introduce the token position information of the CTC model into the training process of transducer. For $\forall n: 1 \leq n \leq U+T$, there is
\begin{equation}
    \begin{aligned}
        &\mathcal{L}_{transducer+fsr}=\mathcal{L}_{transducer}+\lambda\mathcal{L}_{fsr} \\
        &=-\text{ln}\sum_{(t, u):t+u=n}[ P(\mathcal{A}_{t,u}|\bm{x}) \\
        &+ \lambda\mathcal{C}_t^{nb}P(\mathcal{A}^{nb}_{t,u}|\bm{x}) + \lambda\mathcal{C}_t^{b}P(\mathcal{A}^{b}_{t,u}|\bm{x}) ]
    \end{aligned}
\end{equation}
where $\lambda$ is the FSR weight utilized to balance the regularization term and transducer loss. The second term $\mathcal{C}_t^{nb}P(\mathcal{A}^{nb}_{t,u}|\bm{x})$ encourages the transducer model to predict the non-blank tokens (vertical transfer in the Fig.2(a)) at the spike position. The third term $\mathcal{C}_t^{b}P(\mathcal{A}^{b}_{t,u}|\bm{x})$ encourages the transducer model to learn the blank position predicted by the CTC model (horizontal transfer in the Fig.2(a)). The gradients of transducer model with respect to blank and non-blank prediction can be further rewritten as
\begin{equation}
    \begin{aligned}
        \frac{\partial \mathscr{L}}{\partial P(k=y_u|t,u)} &= -\frac{(1+\lambda \mathcal{C}_t^{nb} )\alpha(t,u)\beta(t,u+1)}{P(\bm{y}|\bm{x})} \\
        \frac{\partial \mathscr{L}}{\partial P(k=\varnothing|t,u)} &= -\frac{(1+\lambda \mathcal{C}_t^{b})\alpha(t,u)\beta(t+1,u)}{P(\bm{y}|\bm{x})}
    \end{aligned}
\end{equation}
During the training process, the FSR regularization will make the transducer models tend to choose the possible alignment paths matching with the position prior knowledge from the CTC module. Now that we introduce an external CTC project layer, the final joint loss function can be expressed as:
\begin{equation}
    \mathcal{L}_{joint}=\mathcal{L}_{CTC} + \mathcal{L}_{transducer}+\lambda\mathcal{L}_{fsr}
\end{equation}

\subsection{The Fast-Skip Inference}
The traditional transducer models decode the output sequence frame by frame along the time axis. At each decoding step, the linguistic predictor and joint network can be calculated once respectively, which takes a lot of time and resources. The FSR can assist the transducer model to learn how to predict the blank tokens at the same positions as the CTC model.  This makes the transducer model can predict the blank tokens in advance and skip them.

During the inference, the fast-skip transducer first computes the blank probability $\mathcal{C}_t^{b}$ by the CTC project layer. If the $\mathcal{C}_t^{b}$ is great than the skip-trigger threshold $\delta$, the $t$-frame will be skipped. Otherwise, the $t$-th frame will be marked as triggered and the transducer model will continue to decode based on the current acoustic encoded frame. In order to avoid the deletion error caused by the misalignment of the predicted blank tokens, we set a fixed-length spike-window $(W_{left}, W_{right})$ based on the left and right frames of the triggered position. The $W_{left}$ and $W_{right}$ is the number of the left and right frames. The operation of expanding will sacrifice the inference speed to keep the performance degradation of the model as small as possible.

%\begin{algorithm}[h]
%    \caption{The Fast-Skip Inference Algorithm}
%    \begin{algorithmic}[1]
%        \FOR{each $i \in [1,9]$}
%        \STATE initialize a tree $T_{i}$ with only a leaf (the root);\
%        \STATE $T=T \cup T_{i};$\
%        \ENDFOR
%        \FORALL {$c$ such that $c \in RecentMBatch(E_{n-1})$}
%        \label{code:TrainBase:getc}
%        \STATE $T=T \cup PosSample(c)$;
%        \label{code:TrainBase:pos}
%        \ENDFOR
%        \FOR{$i=1$; $i<n$; $i++$ }
%        \STATE $//$ Your source here;
%        \ENDFOR
%        \FOR{$i=1$ to $n$}
%        \STATE $//$ Your source here;
%        \ENDFOR
%        \STATE $//$ Reusing recent base classifiers.
%        \label{code:recentStart}
%        \WHILE {$(|E_n| \leq L_1 )and( D \neq \phi)$}
%        \STATE Selecting the most recent classifier $c_i$ from $D$;
%        \STATE $D=D-c_i$;
%        \STATE $E_n=E_n+c_i$;
%        \ENDWHILE
%        \label{code:recentEnd}
%    \end{algorithmic}
%\end{algorithm}

\section{Experiments and Results}
\subsection{Dataset}
\vspace{-5pt}
In this work, all experiments are conducted on a public Mandarin speech corpus AISHELL-1\footnote{https://openslr.org/33/}. The training set contains about 150 hours of speech (120,098 utterances) recorded by 340 speakers. The development set contains about 20 hours (14,326 utterances) recorded by 40 speakers. And about 10 hours (7,176 utterances / 36109 seconds) of speech is used as the test set. The speakers of different sets are not overlapped.

\subsection{Experimental Setup}
\vspace{-5pt}
For all experiments, we use 80-dimensional Mel-filter bank coefficients (FBank) features computed on a 25ms window with a 10ms shift. Each feature is re-scaled to have zero mean and unit variance for each speaker. We chose 4233 characters (including a blank token '$\varnothing$', a unknown token \texttt{<UNK>} and a padding token \texttt{<PAD>} ) as model units.

In this paper, we only compared the experiments based on the transformer-transducer(T-T) model \cite{Tian2019}. The T-T consists of 12 encoder blocks and 6 decoder blocks. There are 4 heads in multi-head attention. Both the output size of the multi-head attention and the feed-forward layers are 320. The hidden size of feed-forward layers is 1280. The 2D convolution front end utilizes two-layer time-axis CNN with ReLU activation, stride size 2, channels 320, and kernel size 3. The joint network contains an acoustic project linear layer, a linguistic project linear layer, and an output linear layer with input size 640. We adopt an Adam optimizer with warmup steps 12000 and the learning rate scheduler reported in \cite{vaswani2017attention}. After 150000 epochs, we average the parameters saved in the last 20 checkpoints. We also use the time-mask and frequency-mask method proposed in \cite{park2019specaugment} instead of speed perturbation. We just apply the simple greedy search method for the following experiments. The more results based on other decoding methods will be reported in the future work. Besides, we did not utilize the external language models.

We use the character error rate (CER) to evaluate the performance of different models. For evaluating the inference speed of different models, we decode utterances one by one to compute real-time factor (RTF) on the test set. The RTF is the time taken to decode one second of speech. All experiments are conducted on a GeForce RTX 2080Ti 12G GPU.
\vspace{-5pt}
\subsection{Results}
\vspace{-5pt}
\subsubsection{Comparison of the Models with Different Weight $\lambda$}
\vspace{-10pt}
\begin{table}[h]
    \caption{Comparison of the Models with Different Weight $\lambda$ (CER \%).}
    \vspace{-5pt}
    \centering
    \label{tab:weight}
    \begin{tabular}{|c|c|c|c|}
        \hline
        No. & FSR Weight $\lambda$ & Dev &  Test \\
        \hline
        \hline
        1 & 0.000 & 13.18  & 13.95  \\
        2 & 0.001 & 7.56 & 8.04  \\
        3 & 0.003 & 7.07 & 7.78  \\
        4 & 0.005 & 6.95 & 7.52  \\
        5 & 0.01 &  \textbf{6.80} & \textbf{7.36}  \\
        6 & 0.02 & 6.91 & 7.58  \\
        \hline
    \end{tabular}
    \vspace{-5pt}
\end{table}

\begin{figure*}[t]
    \centering
    \includegraphics[width=0.99\linewidth]{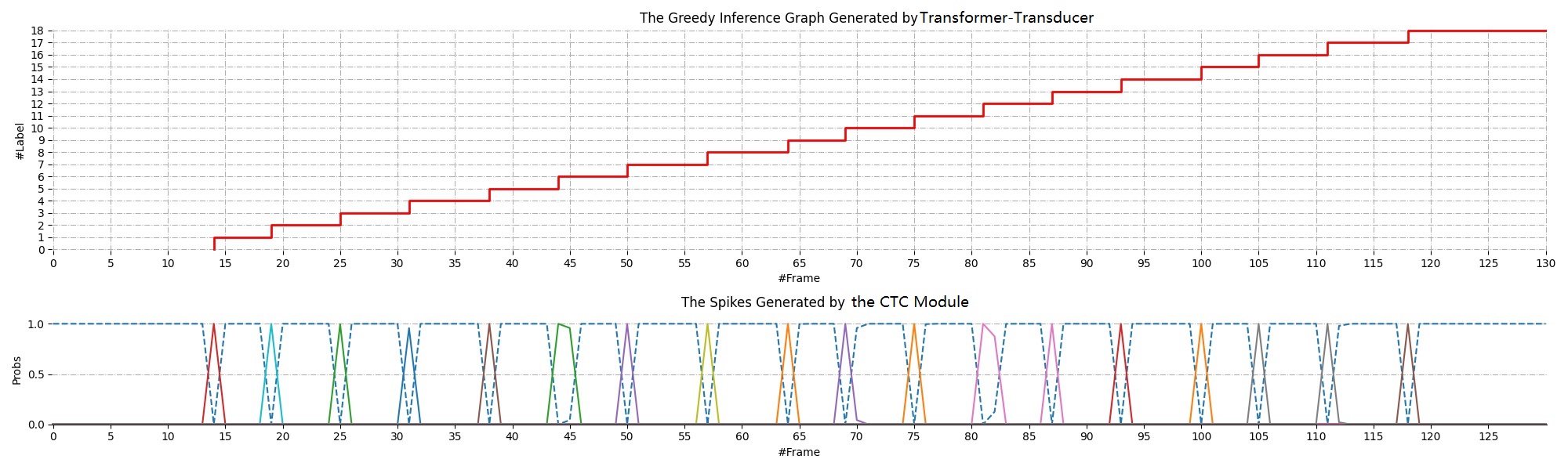}
    \vspace{-10pt}
    \caption{The Visualization of Alignment Relationship between the Transducer Model and CTC Model. The upper figure captures the most possible alignment path of the transformer-transducer during the greedy search. The lower one is the spikes graph predicted by the CTC module. The spikes with different colors stand for different tokens. The blue dot line indicates the blank probability.}
    \label{fig:alignment}
    \vspace{-15pt}
\end{figure*}

In this section, we compare the models trained with different FSR weights $\lambda$. All the  evaluated models set skip-trigger threshold $\delta$ to 0.5 and spike-window to \textit{(1, 1)}. The results show that the model with an FSR weight of 0.01 (No.5) can achieve the best performance on both the development and test set. When the weight $\lambda$ is equal to 0 (No.1), the regularization is abandoned, which is equivalent to utilize a multi-task training method to accelerate the convergence of the transducer model. Under this condition, directly decoding with fast-skip inference may make the model mistakenly skip some key frames and lead to large performance degradation. In the following experiments, we adopt the FSR weight 0.01 as the default parameter.
\vspace{-5pt}
\subsubsection{Comparison of the Length of Expanding Window}
\vspace{-5pt}
Even applying the fast-skip method, we can't guarantee that the transducer model can learn the blank alignment accurately. We construct a fixed-length window around the trigger spike, and all the acoustic frames in the window will be triggered, which could avoid missing some key frames due to the misalignment. As shown in Table.2, the expanding window is indeed able to improve the model performance. When the window width reaches a certain level, the performance of the model is no longer improved, which also confirms that the FSR could make the blank position predicted by these two models close. As the window becomes wider, the computation of the model increases gradually, which leads to the decrease of the inference speed.
\begin{table}[h]
    \caption{Comparison of the Length of Expanding Window (CER \%).}
    \vspace{-10pt}
    \centering
    \label{tab:window}
    \begin{tabular}{|c|c|c|c|c|c|}
        \hline
        No. & $W_{left}$ &  $W_{right}$  & Dev & Test & RTF \\
        \hline
        \hline
        1 & 0 & 0 & 7.03 & 7.75 & 0.0352  \\
        2 & 1 & 0 & 6.92 & 7.50 & 0.0473  \\
        3 & 0 & 1 & 6.90 & 7.49 & 0.0480  \\
        4 & 1 & 1 & 6.80 & 7.36 & 0.0601  \\
        5 & 2 & 1 & 6.64 & 7.21 & 0.0738  \\
        6 & 1 & 2 & 6.64 & 7.20 & 0.0727  \\
        \hline
    \end{tabular}
    \vspace{-10pt}
\end{table}
\vspace{-10pt}
\subsubsection{Ablation Study}
\vspace{-10pt}
\begin{table}[h]
    \caption{Ablation Study on the Inference Process (CER \%).}
    \vspace{-5pt}
    \centering
    \label{tab:ablation}
    \begin{tabular}{|c|c|c|c|c|c|c|}
    \hline
    Model & No. & FSR & FSI & Dev &  Test  & RTF \\
    \hline
    \hline
%    Transformer-Transducer & \CheckmarkBold & \XSolidBrush & 0.0 & 0.0  & 0.0001 \\
%    \quad w/o CTC & \CheckmarkBold & \XSolidBrush & 0.0 & 0.0  & 0.0001 \\
%    \hline
    \multirow{4}*{\shortstack{T-T}} & A & \CheckmarkBold & \CheckmarkBold & 6.80 & 7.36  & 0.0601 \\
    & B & \CheckmarkBold & \XSolidBrush & 6.40 & 7.15  & 0.1983 \\
    & C & \XSolidBrush & \CheckmarkBold & 13.18 & 13.95  & 0.0615 \\
    & D & \XSolidBrush & \XSolidBrush & 6.32 & 7.12  & 0.2122 \\
    % & \XSolidBrush & \XSolidBrush & \XSolidBrush & 0.0 & 0.0  & 0.0001 \\
    \hline
%    \hline
%    RNN-Transducer & \CheckmarkBold & \XSolidBrush & 0.0 & 0.0  & 0.0001 \\
%    \quad w/o CTC & \CheckmarkBold & \XSolidBrush & 0.0 & 0.0  & 0.0001 \\
%    \hline
%    \multirow{4}*{\shortstack{RNN-T}} & \CheckmarkBold & \CheckmarkBold & \CheckmarkBold & 0.0 & 0.0  & 0.0001 \\
%    & \CheckmarkBold & \CheckmarkBold & \XSolidBrush & 0.0 & 0.0  & 0.0001 \\
%    & \CheckmarkBold & \XSolidBrush & \CheckmarkBold & 0.0 & 0.0  & 0.0001 \\
%    & \XSolidBrush & \XSolidBrush & \XSolidBrush & 0.0 & 0.0  & 0.0001 \\
%    \hline
    \end{tabular}
    \vspace{-5pt}
\end{table}
This section focuses on the importance of different components. All the evaluated models with FSI set skip-trigger threshold $\delta$ to 0.5 and spike-window to \textit{(1, 1)}. In order to facilitate the description, we mark the above four models as \texttt{A}, \texttt{B}, \texttt{C} and \texttt{D} respectively. Comparing model \texttt{A} with \texttt{B}, we find that the FSI could improve the inference speed nearly 4 times although he also brings a little performance degradation. The performance degradation can be alleviated by enlarging the window in the inference process. In the comparison of model \texttt{A} and \texttt{C}, it's easy to observe that the transducer model without FSR has a great performance degradation, which also proves that the fast-skip regularization is effective. The model \texttt{D} trained jointly with a CTC module can be regarded as a baseline model. Compared with the model \texttt{D}, the transducer models with FSR (\texttt{A} and \texttt{B}) just have a little performance degradation.
\vspace{-5pt}
\subsubsection{The Visual analysis of Alignment}
\vspace{-5pt}
To further study the impact of the FSR, we draw the picture as shown in Figure.3, which captures the positional alignment relationship between the transducer model and the CTC model. There is a frame-by-frame correspondence between these two subgraphs. We notice that the predicted alignment path in the upper part of Fig.3 starts at the 15-th acoustic frame. And the CTC model predicts a spike based on the same frame. It is no coincidence that each vertical shift of the T-T corresponds to a spike predicted by the CTC module. This proves our assumption that the transducer model could skip the blank position predicted by the CTC model in advance to reduce the computation and accelerate the inference.
\vspace{-5pt}
\section{Conclusions and Future Works}
\vspace{-5pt}
A typical transducer model decodes the output sequence conditioned on the current acoustic state and previously predicted tokens step by step. Statistically, the number of blank tokens in the prediction results accounts for nearly 90\% of all tokens. It takes a lot of computation and time to predict the blank label, but it will not appear in the final output sequence. Consequently, we assume that the inference process of the transducer can be accelerated greatly if the blank tokens can be predicted in advance just depending on simple computation, and then skipped. This paper proposes a method named fast-skip regularization, which tries to align the blank position predicted by a transducer with that predicted by a CTC model. During the inference, the transducer model can skip the frames predicted by a simple CTC project layer as blank tokens. All experiments are conducted on a public Chinese mandarin dataset AISHELL-1. The results show that the fast-skip regularization can indeed help the transducer model learn the blank position predicted by the CTC module. The inference with fast-skip can be speeded up nearly 4 times with only a little performance degradation. In the future, we will focus on how to accelerate the training and inference simultaneously by applying the fast-skip regularization. We also provide more results based on the different inference methods and different dataset.
\newpage
\bibliographystyle{IEEEtran}

\bibliography{mybib}

\begin{thebibliography}{10}
\providecommand{\url}[1]{#1}
\csname url@samestyle\endcsname
\providecommand{\newblock}{\relax}
\providecommand{\bibinfo}[2]{#2}
\providecommand{\BIBentrySTDinterwordspacing}{\spaceskip=0pt\relax}
\providecommand{\BIBentryALTinterwordstretchfactor}{4}
\providecommand{\BIBentryALTinterwordspacing}{\spaceskip=\fontdimen2\font plus
\BIBentryALTinterwordstretchfactor\fontdimen3\font minus
  \fontdimen4\font\relax}
\providecommand{\BIBforeignlanguage}[2]{{%
\expandafter\ifx\csname l@#1\endcsname\relax
\typeout{** WARNING: IEEEtran.bst: No hyphenation pattern has been}%
\typeout{** loaded for the language `#1'. Using the pattern for}%
\typeout{** the default language instead.}%
\else
\language=\csname l@#1\endcsname
\fi
#2}}
\providecommand{\BIBdecl}{\relax}
\BIBdecl

\bibitem{graves2012sequence}
A.~Graves, ``Sequence transduction with recurrent neural networks,''
  \emph{arXiv preprint arXiv:1211.3711}, 2012.

\bibitem{rao2017exploring}
K.~Rao, H.~Sak, and R.~Prabhavalkar, ``Exploring architectures, data and units
  for streaming end-to-end speech recognition with rnn-transducer,'' in
  \emph{2017 IEEE Automatic Speech Recognition and Understanding Workshop
  (ASRU)}.\hskip 1em plus 0.5em minus 0.4em\relax IEEE, 2017, pp. 193--199.

\bibitem{bagby2018efficient}
T.~Bagby, K.~Rao, and K.~C. Sim, ``Efficient implementation of recurrent neural
  network transducer in tensorflow,'' in \emph{2018 IEEE Spoken Language
  Technology Workshop (SLT)}.\hskip 1em plus 0.5em minus 0.4em\relax IEEE,
  2018, pp. 506--512.

\bibitem{sainath2019two}
T.~N. Sainath, R.~Pang, D.~Rybach, Y.~He, R.~Prabhavalkar, W.~Li, M.~Visontai,
  Q.~Liang, T.~Strohman, Y.~Wu \emph{et~al.}, ``Two-pass end-to-end speech
  recognition,'' \emph{arXiv preprint arXiv:1908.10992}, 2019.

\bibitem{li2019improving}
J.~Li, R.~Zhao, H.~Hu, and Y.~Gong, ``Improving rnn transducer modeling for
  end-to-end speech recognition,'' in \emph{2019 IEEE Automatic Speech
  Recognition and Understanding Workshop (ASRU)}.\hskip 1em plus 0.5em minus
  0.4em\relax IEEE, 2019, pp. 114--121.

\bibitem{graves2013speech}
G.~Alex, M.~Abdel-rahman, and H.~Geoffrey, ``Speech recognition with deep
  recurrent neural networks,'' in \emph{2013 IEEE International Conference on
  Acoustics, Speech and Signal Processing}, 2013, pp. 6645--6649.

\bibitem{he2019streaming}
Y.~He, T.~N. Sainath, R.~Prabhavalkar, I.~McGraw, R.~Alvarez, D.~Zhao,
  D.~Rybach, A.~Kannan, Y.~Wu, R.~Pang \emph{et~al.}, ``Streaming end-to-end
  speech recognition for mobile devices,'' in \emph{ICASSP 2019-2019 IEEE
  International Conference on Acoustics, Speech and Signal Processing
  (ICASSP)}.\hskip 1em plus 0.5em minus 0.4em\relax IEEE, 2019, pp. 6381--6385.

\bibitem{Tian2019}
Z.~Tian, J.~Yi, J.~Tao, Y.~Bai, and Z.~Wen, ``{Self-Attention Transducers for
  End-to-End Speech Recognition},'' in \emph{Proc. Interspeech 2019}, 2019, pp.
  4395--4399.

\bibitem{zhang2020transformer}
Q.~Zhang, H.~Lu, H.~Sak, A.~Tripathi, E.~McDermott, S.~Koo, and S.~Kumar,
  ``Transformer transducer: A streamable speech recognition model with
  transformer encoders and rnn-t loss,'' in \emph{ICASSP 2020-2020 IEEE
  International Conference on Acoustics, Speech and Signal Processing
  (ICASSP)}.\hskip 1em plus 0.5em minus 0.4em\relax IEEE, 2020, pp. 7829--7833.

\bibitem{yeh2019transformer}
C.-F. Yeh, J.~Mahadeokar, K.~Kalgaonkar, Y.~Wang, D.~Le, M.~Jain, K.~Schubert,
  C.~Fuegen, and M.~L. Seltzer, ``Transformer-transducer: End-to-end speech
  recognition with self-attention,'' \emph{arXiv preprint arXiv:1910.12977},
  2019.

\bibitem{bahdanau2014neural}
D.~Bahdanau, K.~Cho, and Y.~Bengio, ``Neural machine translation by jointly
  learning to align and translate,'' \emph{arXiv preprint arXiv:1409.0473},
  2014.

\bibitem{chorowski2015attention}
C.~J. K, B.~Dzmitry, S.~Dmitriy, C.~Kyunghyun, and B.~Yoshua, ``Attention-based
  models for speech recognition,'' in \emph{Advances in neural information
  processing systems}, 2015, pp. 577--585.

\bibitem{chan2016listen}
W.~Chan, N.~Jaitly, Q.~Le, and O.~Vinyals, ``Listen, attend and spell: A neural
  network for large vocabulary conversational speech recognition,'' in
  \emph{2016 IEEE International Conference on Acoustics, Speech and Signal
  Processing (ICASSP)}.\hskip 1em plus 0.5em minus 0.4em\relax IEEE, 2016, pp.
  4960--4964.

\bibitem{kim2017joint}
K.~Suyoun, H.~Takaaki, and W.~Shinji, ``Joint ctc-attention based end-to-end
  speech recognition using multi-task learning,'' in \emph{2017 IEEE
  international conference on acoustics, speech and signal processing
  (ICASSP)}.\hskip 1em plus 0.5em minus 0.4em\relax IEEE, 2017, pp. 4835--4839.

\bibitem{dong2018speech}
D.~Linhao, X.~Shuang, and X.~Bo, ``Speech-transformer: a no-recurrence
  sequence-to-sequence model for speech recognition,'' in \emph{2018 IEEE
  International Conference on Acoustics, Speech and Signal Processing
  (ICASSP)}.\hskip 1em plus 0.5em minus 0.4em\relax IEEE, 2018, pp. 5884--5888.

\bibitem{graves2006connectionist}
G.~Alex, F.~Santiago, G.~Faustino, and S.~Jürgen, ``Connectionist temporal
  classification: labelling unsegmented sequence data with recurrent neural
  networks,'' in \emph{Proceedings of the 23rd international conference on
  Machine learning}, 2006, pp. 369--376.

\bibitem{hannun2014deep}
H.~Awni~Y., C.~Carl, C.~Jared, C.~Bryan, D.~Greg, E.~Elsen, P.~Ryan,
  S.~Sanjeev, S.~Shubho, A.~Coates, and N.~Andrew~Y., ``Deep speech: Scaling up
  end-to-end speech recognition,'' \emph{arXiv preprint arXiv:1412.5567}, 2014.

\bibitem{li2019jasper}
L.~Jason, L.~Vitaly, G.~Boris, L.~Ryan, K.~Oleksii, C.~Jonathan~M., N.~Huyen,
  and G.~Ravi~Teja, ``Jasper: An end-to-end convolutional neural acoustic
  model,'' in \emph{Interspeech 2019}, 2019, pp. 71--75.

\bibitem{jaitly2015neural}
N.~Jaitly, D.~Sussillo, Q.~V. Le, O.~Vinyals, I.~Sutskever, and S.~Bengio, ``A
  neural transducer,'' \emph{arXiv preprint arXiv:1511.04868}, 2015.

\bibitem{tian2019synchronous}
Z.~Tian, J.~Yi, Y.~Bai, J.~Tao, S.~Zhang, and Z.~Wen, ``Synchronous
  transformers for end-to-end speech recognition,'' in \emph{ICASSP 2020 IEEE
  International Conference on Acoustics, Speech and Signal Processing
  (ICASSP)}.\hskip 1em plus 0.5em minus 0.4em\relax IEEE, 2020, pp. 5666--5670.

\bibitem{sak2017recurrent}
H.~Sak, M.~Shannon, K.~Rao, and F.~Beaufays, ``Recurrent neural aligner: An
  encoder-decoder neural network model for sequence to sequence mapping.'' in
  \emph{Interspeech}, vol.~8, 2017, pp. 1298--1302.

\bibitem{dong2018extending}
L.~Dong, S.~Zhou, W.~Chen, and B.~Xu, ``Extending recurrent neural aligner for
  streaming end-to-end speech recognition in mandarin,'' \emph{arXiv preprint
  arXiv:1806.06342}, 2018.

\bibitem{dong2019self}
L.~Dong, F.~Wang, and B.~Xu, ``Self-attention aligner: A latency-control
  end-to-end model for asr using self-attention network and chunk-hopping,'' in
  \emph{ICASSP 2019-2019 IEEE International Conference on Acoustics, Speech and
  Signal Processing (ICASSP)}.\hskip 1em plus 0.5em minus 0.4em\relax IEEE,
  2019, pp. 5656--5660.

\bibitem{kim2020accelerating}
J.~Kim and Y.~Lee, ``Accelerating rnn transducer inference via one-step
  constrained beam search,'' \emph{arXiv preprint arXiv:2002.03577}, 2020.

\bibitem{ghodsi2020rnn}
M.~Ghodsi, X.~Liu, J.~Apfel, R.~Cabrera, and E.~Weinstein, ``Rnn-transducer
  with stateless prediction network,'' in \emph{ICASSP 2020-2020 IEEE
  International Conference on Acoustics, Speech and Signal Processing
  (ICASSP)}.\hskip 1em plus 0.5em minus 0.4em\relax IEEE, 2020, pp. 7049--7053.

\bibitem{zhang2021tiny}
Y.~Zhang, S.~Sun, and L.~Ma, ``Tiny transducer: A highly-efficient speech
  recognition model on edge devices,'' \emph{arXiv preprint arXiv:2101.06856},
  2021.

\bibitem{yu2020fastemit}
J.~Yu, C.-C. Chiu, B.~Li, S.-y. Chang, T.~N. Sainath, Y.~He, A.~Narayanan,
  W.~Han, A.~Gulati, Y.~Wu \emph{et~al.}, ``Fastemit: Low-latency streaming asr
  with sequence-level emission regularization,'' \emph{arXiv preprint
  arXiv:2010.11148}, 2020.

\bibitem{li2020better}
B.~Li, A.~Gulati, J.~Yu, T.~N. Sainath, C.-C. Chiu, A.~Narayanan, S.-Y. Chang,
  R.~Pang, Y.~He, J.~Qin \emph{et~al.}, ``A better and faster end-to-end model
  for streaming asr,'' \emph{arXiv preprint arXiv:2011.10798}, 2020.

\bibitem{narayanan2020cascaded}
A.~Narayanan, T.~N. Sainath, R.~Pang, J.~Yu, C.-C. Chiu, R.~Prabhavalkar,
  E.~Variani, and T.~Strohman, ``Cascaded encoders for unifying streaming and
  non-streaming asr,'' \emph{arXiv preprint arXiv:2010.14606}, 2020.

\bibitem{vaswani2017attention}
A.~Vaswani, N.~Shazeer, N.~Parmar, J.~Uszkoreit, L.~Jones, A.~N. Gomez,
  {\L}.~Kaiser, and I.~Polosukhin, ``Attention is all you need,'' in
  \emph{Advances in neural information processing systems}, 2017, pp.
  5998--6008.

\bibitem{park2019specaugment}
D.~S. Park, W.~Chan, Y.~Zhang, C.-C. Chiu, B.~Zoph, E.~D. Cubuk, and Q.~V. Le,
  ``Specaugment: A simple data augmentation method for automatic speech
  recognition,'' \emph{arXiv preprint arXiv:1904.08779}, 2019.

\end{thebibliography}

\end{document}